\documentclass[]{spie}  

\usepackage{amsmath,amsfonts,amssymb}
\usepackage{graphicx}
\usepackage[colorlinks=true, allcolors=blue]{hyperref}
\usepackage{color}

\title{High-contrast spectroscopy testbed for Segmented Telescopes: instrument overview and development progress}

\author{N.~Jovanovic$^{a}$, G.~Ruane$^{a}$, D.~Echeverri$^{a}$, J.R.~Delorme$^{a}$, D.~Mawet$^{a,b}$, J.~Fucik$^{a}$, J.K.~Wallace$^{b}$,  C. Coker$^{b}$, A. Delacroix$^{a}$, N. Levraud$^{a}$, J.D.~Llop~Sayson$^{a}$, J.~Wang$^{a}$, R.~Riddle$^{a}$, M. A. Millar-Blanchaer$^{b}$

$^{a}$ Department of Astronomy, California Institute of Technology, 1200 E. California Blvd., Pasadena, CA, USA 91125; \\
$^{b}$ Jet Propulsion Laboratory, California Institute of Technology, 4800 Oak Grove Drive, Pasadena, CA, USA 91109;
}

\authorinfo{Further author information: (Send correspondence to Nemanja Jovanovic)\\Nemanja Jovanovic: E-mail:  jovanovic.nem@gmail.com, Telephone: +1 626 395 1214}

\begin{document} 
\maketitle

\begin{abstract}
The High Contrast spectroscopy testbed for Segmented Telescopes (HCST) is being developed at Caltech. It aims at addressing the technology gap for future exoplanet imagers and providing the U.S. community with an academic facility to test components and techniques for high contrast imaging, focusing on segmented apertures proposed for future ground-based (TMT, ELT) and space-based telescopes (HabEx, LUVOIR). 

We present an overview of the design of the instrument and a detailed look at the testbed build and initial alignment. We offer insights into stumbling blocks encountered along the path and show that the testbed is now operational and open for business. We aim to use the testbed in the future for testing of high contrast imaging techniques and technologies with amongst with thing, a TMT-like pupil. 
\end{abstract}


\section{INTRODUCTION}
\label{sec:intro}  
Direct imaging is a crucial area of exoplanetory science that can be used to constrain the luminosity, orbit, chemical composition and abundances of exoplanets. The difficulty however, is being able to see the faint exoplanet in the vicinity of the much brighter host star. In practice, the atmosphere spreads the photons of the star over a large area in the focal plane complicating things even further. Adaptive Optics (AO) systems are typically used to try to restore the point spread function (PSF), but even extreme AO (ExAO) systems that reach $80$--$90\%$ Strehl ratios in the H-band, like GPI~\cite{macintosh2014}, SPHERE~\cite{beuzit2008} and SCExAO~\cite{jovanovic2015}, aren’t perfect leaving some residual light in the form of fluctuating speckles in and around the PSF of the star. In addition, the quasi-static speckles arising from diffraction from the telescope aperture, the spiders, and other optical aberrations also add bright features to the focal plane and create a sea of speckles that limit the detectability of faint companions close to the host star. For these reasons, all substellar companions imaged to data are large (multiple Jupiter mass) and beyond 5 AU from their host star (out to 1000's of AU). 

The technique of direct imaging would be extremely valuable if it could be applied to study exoplanets at lower masses and at smaller separations ($<5$~AU), which are more representative of our inner solar system. To do this though, significant advances need to be made to high contrast imaging techniques. For this reason, the High Contrast testbed for Segmented Telescopes (HCST) is being developed in the Exoplanet Technology Laboratory at Caltech. The aim of the testbed is to 
\begin{itemize}
	\item {Develop advanced coronagraphs. This will mostly focus on the development of vortex coronagraphs with different charges combined with various upstream pupil apodization schemes for extreme contrasts ($10^{-8}$).}
	\item {Utilize common-path wavefront sensing. To reach the extreme contrasts the HCST testbed will utilize various forms of common path wavefront sensing such as speckle nulling~\cite{borde2006} and electric field conjugation (EFC)~\cite{giveon2011} for example.}
	\item{Test on segmented apertures. The HCST will aim at showing these extreme contrasts on segmented apertures which represent the future giant segmented mirror telescopes (GSMTs) and space telescopes such as HabEx and LUVOIR. The goal is to show that phasing and amplitude errors can be controlled on such apertures at the contrasts required.}
	\item{Demonstrate high dispersion coronagraphy. High dispersion coronagraphy (HDC) relies on using a single mode fiber in the focal plane downstream of the coronagraph to send the known planets light to a high resolution spectrograph to study its properties~\cite{wang2017}. The technique is in its infancy and needs substantial development. One aspect that the HCST will focus on is wavefront control and star light suppression through the single mode fiber~\cite{mawet2017}. This is a potential avenue to achieving extremely high contrasts.}
\end{itemize}

In this paper we outline the HCST, report the status of the testbed and offer insights into areas of interest the testbed will focus on in the coming years.

\section{The High Contrast testbed for Segmented Telescopes}~\label{sec:HCST}
\begin{figure} [b!]
   \begin{center}
   \begin{tabular}{c} 
   \includegraphics[width=0.98\textwidth]{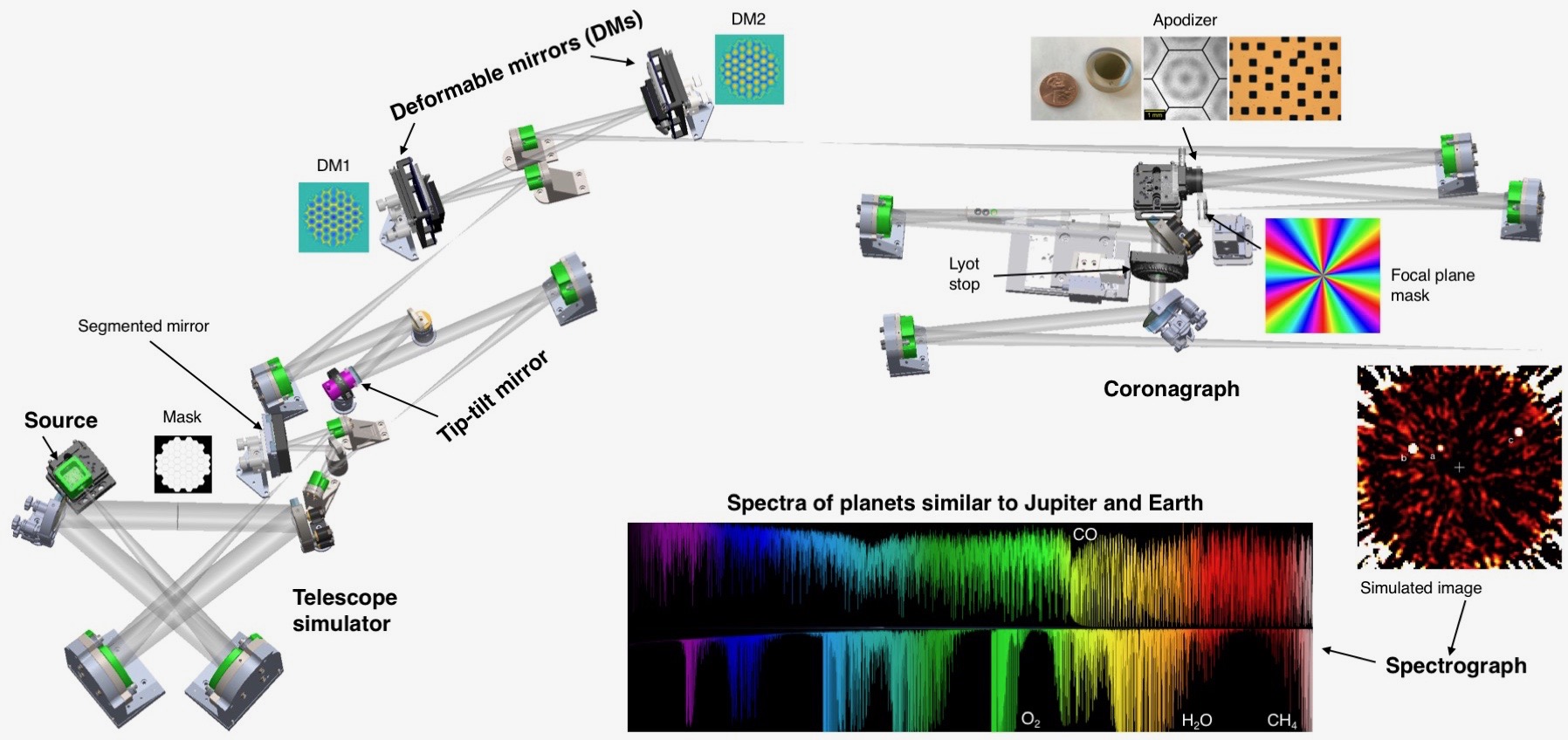}
   \end{tabular}
   \end{center}
   \caption[example] 
   { \label{fig:schematic} 
A schematic of the HCST testbed.}
\end{figure}

A schematic of the HCST is shown in Figure~\ref{fig:schematic}.
The system is based on entirely reflective optics to maximize its wavelength range of operation in order to meet the varied science cases of future telescopes. The first pair of off-axis parabolas (OAPs) reimage the beam from the output of a single-mode fiber which represents the star. A pupil mask can be used in the interim pupil plane to simulate a segmented telescope aperture or the input pupil mask for a coronagraph. The beam is then collimated and the pupil is projected onto a segmented deformable mirror. This mirror can be used for simulating the correction of phasing errors from segmented telescopes and their subsequent correction. The pupil is then reimaged with a pair of OAPs so that it coincides with a tip/tilt mirror. This mirror will be used to control slow drifts in the bench and fast vibrations. The pupil is once again reimaged with a pair of OAPs onto a Boston Micromachines Corporation (BMC) 1k deformable mirror (DM). This device offers 34 actuators across the pupil with a 3.5 $\mu$m stroke and can operate at kHz speeds. A second similar DM will eventually be placed about 250 mm downstream of the first in a bid to control both amplitude and phase in the system. Another pair of OAPs reimages the pupil once more onto a coronagraph apodizer. These will be developed based on microdot technology~\cite{zhang2018}. The apodizer reflects the part of the beam you want to keep and transmits the part that we don’t want. In the downstream focal plane, the coronagraphic focal plane masks can be deployed. These include a vortex mask for coronagraphy as well as a phase dimple (a cylindrical pit in a transparent substrate) that can be used to realize a Zernike wavefront sensor. This wavefront sensor will be used to correct for the aberrations in the system up to the focal plane mask to ensure maximum light suppression by the coronagraph. A Lyot stop is located in the downstream pupil plane. An opaque laser cut mask will block the unwanted light in the pupil while a reflective ring mirror placed in front of the mask will direct all of the unwanted light outside the geometric pupil towards a low order wavefront sensor (LOWFS) that will be built to control low order aberrations at the coronrgaphic focal plane in real time~\cite{singh2014}. The light transmitted at the Lyot stop is focused onto a science camera (Andor, NEO). Before the focus, a lens can be inserted to image the pupil in the system. There are plans to pick the light off pre-focal plane and inject it into a single mode fiber to advance the HDC technique.

To reach extreme contrasts, the following wavefront control scheme will be used. Firstly, the common-path errors will be minimized up to the focal plane, by the Zernike wavefront sensor. The vortex mask will then be inserted suppressing most of the on-axis light. The LOWFS will be used to control the low order aberrations at the coronagraphic focal plane mask. Tip/tilt will be controlled by the upstream tip/tilt mirror at high speed to eliminate motion due to vibrations and drifts while other low order modes will be controlled by the 1k DM. Once the PSF has been stabilized, speckle nulling and EFC will be used to dig dark holes around the PSF in initially monochromatic and eventually polychromatic light. This approach will allow the HCST to develop coronagraphs and common-path wavefront sensing approaches for extremely high contrast free from atmospheric perturbations. However, there are plans to deploy a turbulence simulator plate adjacent to the input pupil mask which can be used to inject realistic turbulence into the testbed. In addition, a shack-Hartmann wavefront sensor will pick the light off just after the tip/tilt mirror to enable open-loop sensing at high speeds. This will be the primary form of wavefront sensing to be used with turbulence and both the LOWFS and the speckle control in the focal plane will be used as the after burner to this. 

\section{The HCST development path to date}~\label{sec:development}
\begin{figure} [t!]
   \begin{center}
   \begin{tabular}{c} 
   \includegraphics[width=0.6\textwidth]{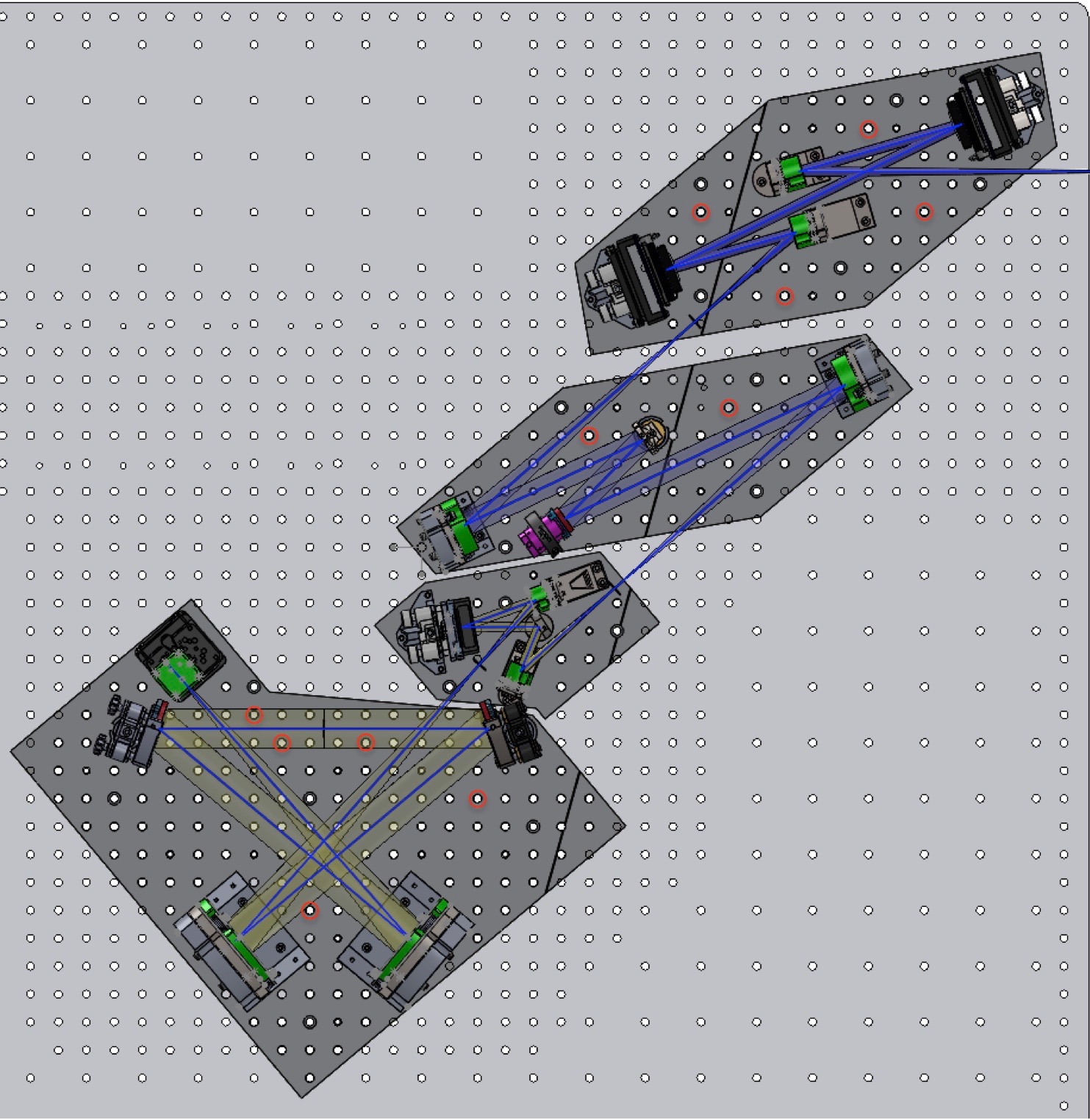}
   \end{tabular}
   \end{center}
   \caption[example] 
   { \label{fig:CAD} 
A CAD of the new baseplate design based on islands. The first four baseplates are shown in this image with the coronagraph baseplate omitted.}
\end{figure}

Integration of the HCST started in late 2017. The original concept was to have a very precise optical and mechanical design of the testbed and hence exploit a single baseplate that had custom holes to locate and secure each optic. Although the CAD model was very precise, a last-minute modification to the optical train did not propagate down to the baseplate design and issues with manufacturing the $5$ mm thick baseplate which spanned $1200$ by $1200$~mm meant that the system could not be aligned on this, after several failed attempts. Since the mounts were custom machined to match the hole patterns on the baseplate, it was not possible to eliminate the baseplate entirely without losing out on substantial work so instead a modular approach to the baseplates was used instead. Five smaller baseplates known as islands were used in the second generation design. Since the baseplates were much smaller, precision polished carbon steel plates which offered high flatness were used and could be easily machined. In this concept, it is possible to 
\begin{itemize}
	\item {Align each plate in isolation before integrating them together}
	\item {Remove islands and replace them as needed. The other islands can be slid around to meet the new beam position if needed}
\end{itemize} 
A CAD drawing showing the first four island baseplates is shown in Fig.~\ref{fig:CAD}. Although each baseplate shape looks complex, they are all actually rectangles with one or two corners removed. In addition, they are designed so that their faces are parallel to each other so that fixed width spacers can be used to align each plate with the next as a starting point. Scribes were implemented to show the beam position on all plates, as well as the location of the focal and pupil planes in an effort to simplify alignment.  

Three techniques were used to align the testbed which included: a Zygo interferometer, a HASO shack-Hartmann wavefront sensor and direct imaging of the beam in various planes. The first two tools provide the wavefront of the beam after propagating through the testbed. The control software for these offers the ability to decompose the wavefront into Zernike modes so that the amplitude of typical aberrations, like astigmatism can be quantified. The third technique relies on the formation of an image and the subsequent inspection of the resulting PSF. When aligning OAPs, astigmatism is the most common aberration and is easily visible in the PSF as the break-up of the first and subsequent Airy rings. By using the shape of the PSF as a proxy, it is possible to tune out the aberrations in the optics.

To promote the high contrast science case, the OAPs were given very tight optical specifications on the wavefront error allowable as shown in table~\ref{tab:OAPspecs}. As seen in the table, tighter tolerances were placed on the figure error for higher spatial frequencies ($>15$ cycles/aperture). This was because the 1k DM has a cutoff frequency of $17$ cycles/aperture and hence can not be used to correct for these modes if they are introduced by the optics. In addition to the tight optical requirements on the OAP surface figures, the beam on each OAP was no larger than $80\%$ of the OAP diameter, meaning the beam subtended an area smaller than the full optic minimizing the wavefront error experienced upon reflection.  

Besides the figure of the OAP, the angle of the OAP with respect to the incoming beam as well as the mounting scheme will also effect the total wavefront error. And given that there are 19 surfaces from the source fiber to the focal plane where the coronagraphic mask will be placed, each optic had to be carefully optimized and characterized to ensure aberrations were minimized throughput the bench. The first step involved mounting each optic in its mount, adjusting the clocking in the case of OAPs, and then adjusting the tension on the locking screws very carefully while monitoring on the Zygo interferometer. A reflective silicon nitride ball was used to reflect the beam from the focus of the OAPs back to the interferometer for interrogation. Through this process, it was determined that each type of OAP behaved very differently. For example OAPs that were thicker compared to their width, were less susceptible to deformation when clamping while OAP one and two in the chain, which were the largest and relatively thin were very sensitive and would deform as soon as the locking mechanism made contact with the optic. A full discussion of our observations can be found in Echeverri et al 2018~\cite{dan2018}. 

\begin{table}[ht]
\caption{A list of the tolerances and specifications requested and met by the manufacturer - Nu-Tek. $\phi$ is the aperture size.} 
\label{tab:OAPspecs}
\begin{center}       
\begin{tabular}{|l|c|c|} 
\hline
\rule[-1ex]{0pt}{3.5ex}  \textbf{Attribute} & \textbf{Description} & \textbf{Specification}\\
\hline
\hline
\rule[-1ex]{0pt}{3.5ex}  Low frequency figure error     & $<3$ cycles/$\phi$, RMS  & $<5$~nm  \\
\hline
\rule[-1ex]{0pt}{3.5ex}  Mid frequency figure error     & 3--15 cycles/$\phi$, RMS  & $<2$~nm  \\
\hline
\rule[-1ex]{0pt}{3.5ex}  High frequency figure error    & $>15$ cycles/$\phi$, RMS  & $<1$~nm  \\
\hline
\rule[-1ex]{0pt}{3.5ex}  Total surface error                 & Over $\phi$, PV                  & $<30$~nm  \\
\hline
\rule[-1ex]{0pt}{3.5ex}  Focal length                       & $\%$ of focal length & $0.1\%$ \\
\hline
\rule[-1ex]{0pt}{3.5ex}  Material                           & Zerodur           & Class 0/MIL-G-174\\
\hline
\end{tabular}
\end{center}
\end{table} 

\begin{figure} [t!]
   \begin{center}
   \begin{tabular}{c} 
   \includegraphics[width=0.98\textwidth]{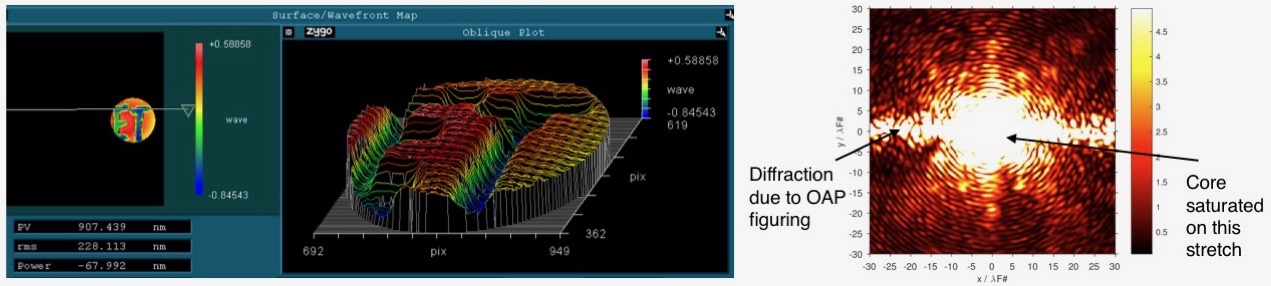}
   \end{tabular}
   \end{center}
   \caption[example] 
   {\label {fig:map} (Left) A map applied to the DM and observed on the zygo. (Right) a saturated PSF with the vortex coronagraph in the beam with a Lyot stop with the flat mirror prior to inserting the DM. Diffration from the OAP figuring can be seen.}
\end{figure}

With the clocking fixed and the mounting induced aberrations minimized, the optics were inserted into the bench one at a time. Each optic had 3 degrees of freedom in the form of actuation to fine adjust its piston and tip/tilt alignment. The Zygo beam was passed through a one inch iris and injected into the testbed at the Lyot plane propagating towards the source. Each optic was placed in the beam path one at a time and its angle of incidence optimized to minimize aberrations including defocus when optimizing the collimation of a beam for example. In places where the beam was collimated, a highly flat mirror was used to reflect the beam for analysis while at focal planes, the silicon nitride ball was used as mentioned earlier. Flat mirrors were initially used at the location of the 2 DMs and the four optics surrounding the segmented DM, were skipped altogether. This decision was made to expedite commissioning as the segmented DM will not be used in the first wave of tests. The process was difficult because the baseplates limited the degrees of freedom available in alignment (i.e. the position of the mounts) and so in order to minimize the total wavefront error, many hours were spent on each optic. One of the flat mirrors was replaced after the entire bench was aligned using the Zygo with an actual 1k DM from BMC. The total wavefront error from the fiber source to the Lyot plane ($20+$ optics), including several flat mirrors that steered the Zygo beam into the bench was $<30$~nm RMS. This was with the native flat map provided by BMC which was not fully optimized. An image showing our ability to control the 1k DM is shown in the left panel of Fig.~\ref{fig:map}. The image shows the letters ET imprinted on the 1k DM, which stands for the Exoplanet Technology Lab. as detected by the Zygo interferometer.The right hand panel shows a coronagraphic PSF of the aligned testbed before the DM was inserted. The PSF is plotted on a linear scale and is saturated to reveal the speckle field. A bright diffractive feature in the horizontal direction can be seen across the middle of the image corresponding to diffraction from the figuring of the OAPs due to the diamond turning process.  

\begin{figure} [t!]
   \begin{center}
   \begin{tabular}{c} 
   \includegraphics[width=0.98\textwidth]{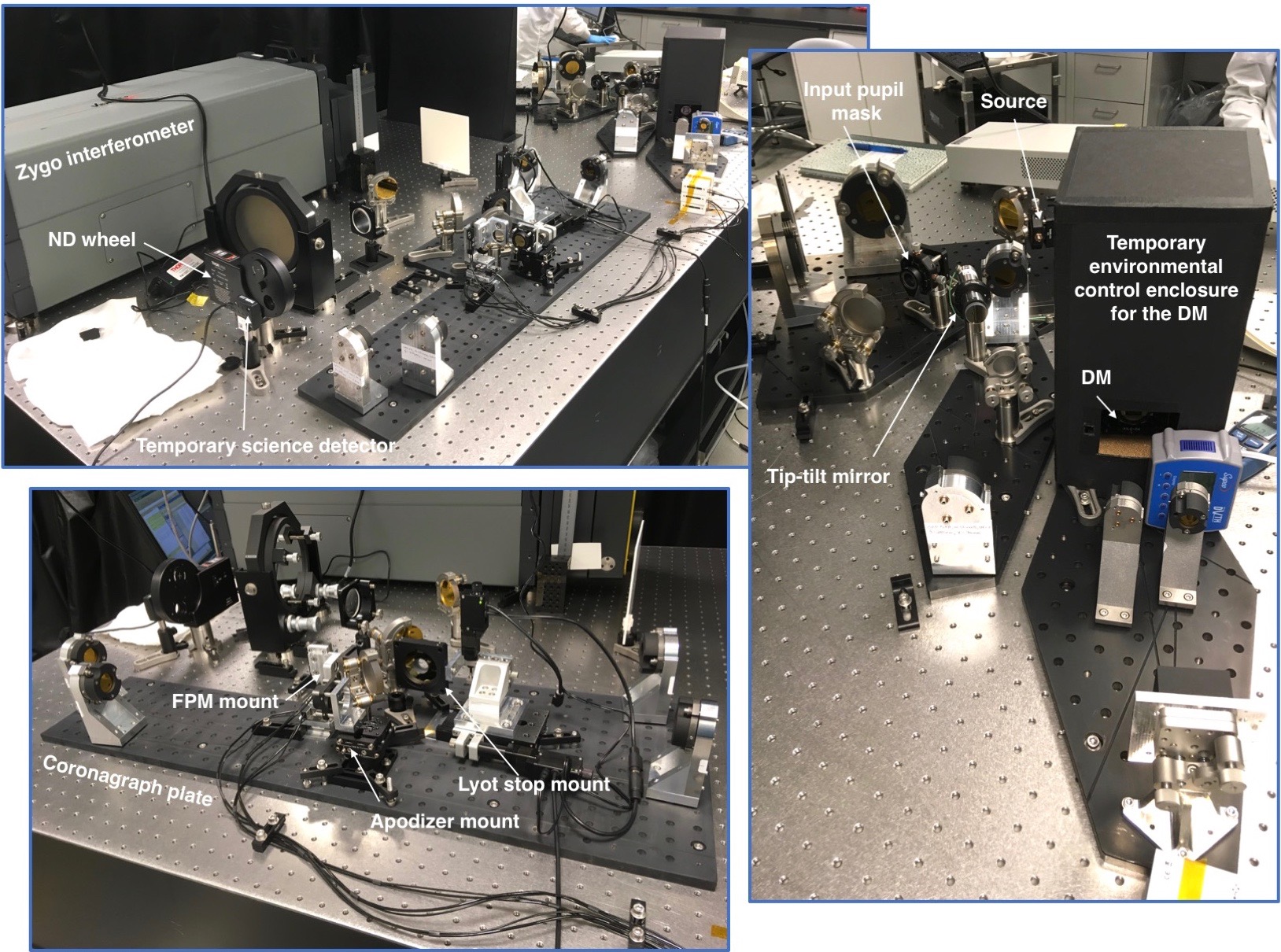}
   \end{tabular}
   \end{center}
   \caption[example] 
   {\label {fig:HCST} The 3 panels show different angles of the recently completed HCST testbed. The island baseplates can clearly be seen along with the Zygo interferometer used to align the instrument. FPM - focal plane mask.}
\end{figure}

Images of the testbed immediately after initial alignment was completed are shown in Fig.~\ref{fig:HCST}. 
The images show the island baseplates and the mounted and aligned optics. Without the enclosure completed at first light, a make shift enclosure was built for the 1k DM, and was purged with dry air to ensure that the DM was operated in a low humidity environment at all times. The Zygo interferometer and some of the large flats used for the initial alignment can also be seen on the bench. 

With the system operational, speckle nulling was implemented with the vortex coronagraph successfully to achieve a contrast below $10^{-7}$ in monochromatic light. For a full discussion of the preliminary performance of the HCST and this result, please see Ruane et al. 2018~\cite{ruane2018}.

\section{Facilitating high contrast}~\label{sec:enclosure}
In order to achieve the extreme contrasts that the HCST is aiming for, environmental controls will be critical. The Exoplanet Technology Laboratory has an exquisite thermal control system that has a peak-to-peak fluctuation of $<1^{\circ}$C over the course of three weeks. The RMS fluctuation was $0.26^{\circ}$C which forms the perfect base to conduct high contrast experiments. 

\begin{figure} [t!]
   \begin{center}
   \begin{tabular}{c} 
   \includegraphics[width=0.98\textwidth]{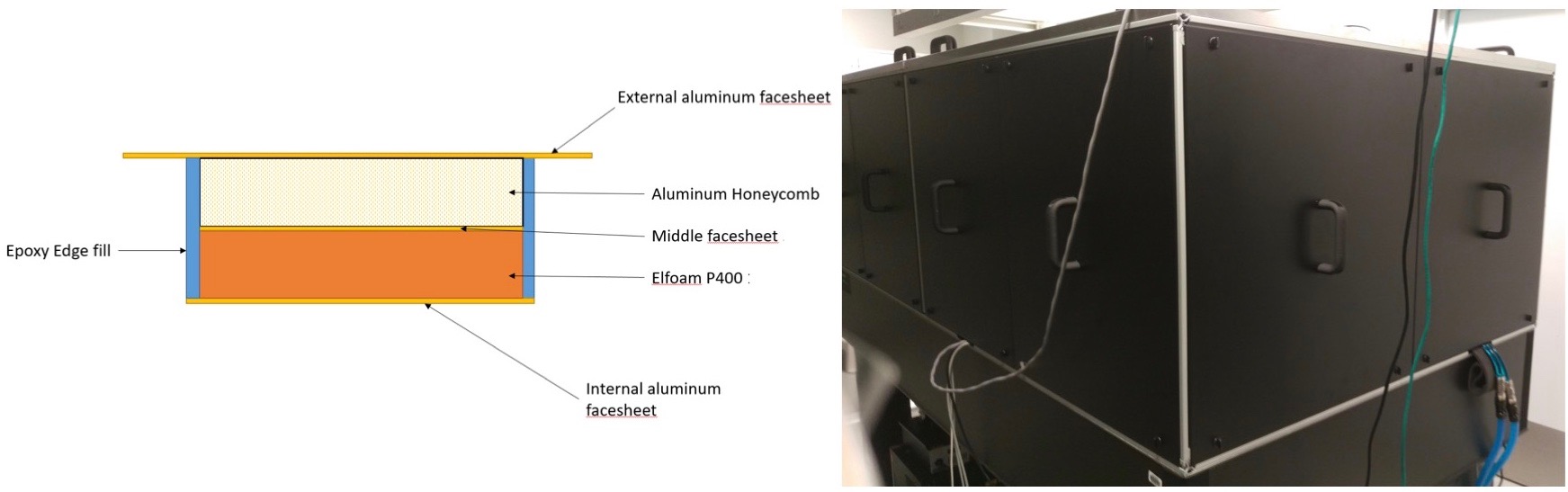}
   \end{tabular}
   \end{center}
   \caption[example] 
   {\label {fig:enclosure} The 3 panels show different angles of the recently completed HCST testbed. The island baseplates can clearly be seen along with the Zygo interferometer used to align the instrument. FPM - focal plane mask.}
\end{figure}

The HCST testbed is further enclosed in a two layer enclosure. The outer enclosure serves two main purposes, namely, to minimize air currents flowing across the bench and improve the thermal isolation of the testbed from the surrounding laboratory. A large outer enclosure design was chosen to all easy access for upgrades and experimental work. Large panels however are prone to acoustic vibrations from external air currents or other laboratory vibrations. To provide both thermal insulation and mechanical stiffness, dual layer panels were designed which had a $0.5$ inch thick foam layer and a $0.5$ inch thick layer of aluminum honeycomb. A cross-sectional schematic of this can be seen in the left panel of Fig.~\ref{fig:enclosure}. The completed outer enclosure is shown in the right hand panel.   

As the turbulence is a function of the volume of an enclosure, a second smaller enclosure will be built inside the outer enclosure that will be only a little larger than the foot print and height of the opto-mechanics. Finally, the OAPs themselves were made from Zerodur, a ceramic material with a near zero coefficient of thermal expansion to further minimize effects of thermal expansion. The dual enclosures, thermally insulating panels, the zerodur OAPs, and the high temperature stability of the room will provide the perfect platform to conduct the extreme contrast experiments planned on the HCST. 

\section{Summary}~\label{sec:summary}
The HCST testbed has been aligned and is now operational. The testbed will focus on developing and demonstrating high contrast imaging techniques and technologies which will be critical to future instruments on GSMTs and space missions. The testbed aims at achieving contrasts of the order of $10^{-8}$ in polychromatic light which is supported by the careful environmental controls put in place around the testbed. The HCST aims at pioneering the HDC technique and especially investigating speckle control through a single mode fiber to enhance contrast and enable the collection of high resolution spectra of terrestrial exoplanets.

\acknowledgments 
 
The authors would like to acknowledge the financial support of the Heising-Simons foundation. G. Ruane is supported by an NSF Astronomy and Astrophysics Postdoctoral Fellowship under award AST-1602444. This work was also supported by the Exoplanet Exploration Program (ExEP), Jet Propulsion Laboratory, California Institute of Technology, under contract to NASA.  

\bibliography{report} 
\bibliographystyle{spiebib} 

\end{document}